\documentclass[12pt]{article}
\usepackage{amsmath}
\usepackage{amstext}
\usepackage{amssymb}
\usepackage{overpic}

\usepackage{upref}  

\usepackage{indentfirst}
\usepackage{float}
\usepackage[mathcal]{euscript}

\usepackage{version}

 \newtheorem{thm}{Theorem}

\newtheorem{conj}[thm]{Conjecture}
\newtheorem{cor}[thm]{Corollary}
\newtheorem{lem}[thm]{Lemma}

\newtheorem{rmk}[thm]{Remark}

\begin{document}

\title{On critical behavior in nonlinear evolutionary PDEs with small viscosity}

\author{B.~Dubrovin\thanks{SISSA, Via Bonomea 265, I-34136 Trieste, Italy, {\tt
dubrovin@sissa.it} and Laboratory of Geometric Methods in Mathematical
Physics, Moscow State University `M.V.Lomonosov',} \and M.~Elaeva\thanks{Laboratory of Geometric Methods in Mathematical
Physics, Moscow State University `M.V.Lomonosov', {\tt mselaeva@gmail.com}}}

\maketitle

\begin{abstract} We address the problem of general dissipative regularization of the quasilinear transport equation. We argue that the local behavior of solutions to the regularized equation near the point of gradient catastrophe for the transport equation is described by the logarithmic derivative of the Pearcey function, a statement generalizing the result of A.M.Il'in \cite{ilin}. We provide some analytic arguments supporting such conjecture and test it numerically.
\end{abstract}

\section*{Introduction}

In this article we address the problem of shock formation in a general dissipative regularization
\begin{equation}\label{regu}
u_t+a(u) u_x = \varepsilon \left[ b(u) u_{xx} +c(u) u_x^2\right]+\varepsilon^2 \left[ b_1(u) u_{xxx} + c_1(u) u_{xx}u_x +d_1(u) u_x^3\right]+\dots
\end{equation}
of the quasilinear transport equation
\begin{equation}\label{tr1}
u_t+a(u)u_x=0, \quad a'(u)\neq 0,\quad u,x\in\mathbb{R}.
\end{equation}
Here $\varepsilon$ is a small positive parameter, the coefficient $b(u)$ does not vanish. In such a study we were inspired by the Universality Conjecture of \cite{univer1} concerning the universal shape of {\it dispersive} shock waves at the point of phase transition from regular to oscillatory behavior. This universal dispersive shock profile is described in terms of a particular solution of certain generalization of the Painlev\'e-I equation (importance of this particular solution in 2D quantum gravity and the theory of Korteweg--de Vries equation was also observed in \cite{moore, bmp, sul}). The universality conjecture for solutions to the Korteweg--de Vries equation with analytic initial data was proved in \cite{cg}. Further numerical evidences supporting the universality conjecture of \cite{univer1} can be found in \cite{dgk}. Another starting point for the present research was the remarkable result by A.M.Il'in (see the book \cite{ilin} and references therein) describing the asymptotics of the generic solution to the equation
\begin{equation}\label{ili}
u_t+a(u) u_x =\varepsilon\, u_{xx}
\end{equation}
at the point of shock formation in terms of the logarithmic derivative of the so-called Pearcey integral (see below the precise formulation of the Il'in's asymptotic formula). In both dispersive and dissipative cases the leading term of the asymptotic formula essentially depends, up to few constants, neither on the choice of a particular generic solution nor on the choice of a particular generic perturbation.

Our main goal is to generalize the Il'in's universality result from the equations \eqref{ili} to the more general case\footnote{It was shown in \cite{ks} that the Il'in formula also works for certain dissipative perturbations of the shallow water equations.} of equations of the form \eqref{regu}. In the present paper we present the conjectural form of such a generalization and describe results of numerical experiments supporting its validity.

The paper is organized as follows. In the first section we explain simple arguments suggesting that, for sufficiently small $\varepsilon$  solutions to the perturbed equation \eqref{regu} can be approximated by solutions to the nonlinear transport equation \eqref{tr1} up to the time of gradient catastrophe of the latter. In order to save the space we omit the terms of order $\epsilon^2$ and higher from the formulae; their contribution to the asymptotic expansions will be of higher order anyway. We then proceed to the precise formulation of the dissipative universality conjecture (see Conjecture \ref{con3} below) describing the leading term of the asymptotic expansion at the point of shock formation. We also give heuristic motivations of this main conjecture. In the last section we present results of numerical experiments supporting the main conjecture. To this end we begin with the standard Burgers equation in order to test the numerical codes based on the finite element analysis. Then we proceed to a particular case of generalized Burgers equation  comparing the numerical solution with the asymptotic formula.

{\bf Acknowledgments}.

This work is
partially supported by the European Research Council Advanced Grant FroM-PDE, by the Russian Federation Government Grant No. 2010-220-01-077 and by PRIN 2008 Grant ``Geometric methods in the theory of nonlinear waves and their applications'' of Italian Ministry of Universities and Researches. Authors thank A.M.Il'in for stimulating discussions.

\vspace{\baselineskip}
\section*{1. Critical behavior in the generalized Burgers equation}\par

Consider the following class of nonlinear PDEs depending on a small parameter $\varepsilon >0$
\begin{equation}\label{burg1}
u_t+a(u) u_x = \varepsilon \left[ b(u) u_{xx} +c(u) u_x^2\right].
\end{equation}
The coefficients $a(u)$, $b(u)$, $c(u)$ are smooth functions, $a'(u)\neq 0$. The class of equations is invariant with respect to arbitrary changes of the dependent variable
$$
u\mapsto \tilde u = f(u), \quad f'(u)\neq 0.
$$
Using such transformations one can reduce \eqref{burg1} to one of the two normal forms
\begin{equation}\label{burg11}
u_t+u\, u_x = \varepsilon \left[ b(u) u_{xx} +c(u) u_x^2\right]
\end{equation}
or
\begin{equation}\label{burg12}
u_t+a(u) u_x=\varepsilon\, b(u) u_{xx}.
\end{equation}

We will study solutions $u=u(x,t; \varepsilon)$ to the Cauchy problem
\begin{equation}\label{burg2}
u(x,0;\varepsilon)=F(x)
\end{equation}
with $\varepsilon$-independent smooth initial data. In the particular case $b(u)=1$, $c(u)=0$ one arrives at the generalized Burgers equation
\begin{equation} \label{ilin1}
u_t+a(u) u_x =\varepsilon\, u_{xx}
\end{equation}
thoroughly studied by A.M.Il'in (see the book \cite{ilin} and references therein). Let us briefly summarize the main results of \cite{ilin}.

For simplicity let us assume the initial data $F(x)$ to be a monotone function on the entire real line $x\in\mathbb R$. The first issue is the comparison of the solution $u(x,t;\varepsilon)$ to the Cauchy problem \eqref{burg1}, \eqref{burg2} with the solution $v=v(x,t)$ to the inviscid equation obtained by setting $\varepsilon$ to $0$ with the \emph{same} initial data
\begin{eqnarray}\label{burg0}
&&
v_t+a(v)v_x=0
\nonumber\\
&&
\\
&&
v(x,0)=F(x).
\nonumber
\end{eqnarray}
The two solutions asymptotically coincide on finite intervals of the $x$-axis for sufficiently small time
$$
|u(x,t;\varepsilon)-v(x,t)| \to 0 \quad \mbox{for}\quad\varepsilon\to 0+, \quad x\in [x_1, x_2], \quad 0\leq t \leq t_1.
$$
However, the lifespan of the solution $v(x,t)$ is finite, due to nonlinear steepening, if the function $a\left(F(x)\right)$ is monotone decreasing on some interval of real axis. In this case the solution to the inviscid equation is defined only on the interval $[0, t_0]$ where
\begin{equation}\label{t0}
t_0=\min_{x\in\mathbb R} \left( -\frac1{[a\left(F(x)\right)]_x}\right).
\end{equation}
Assuming the minimum in \eqref{t0} attained at an isolated point $x=x_0$ to be non-degenerate one arrives at a point of \emph{gradient catastrophe} of the solution $v(x,t)$, i.e., the limit
\begin{equation}\label{v0}
\lim_{t\to t_0, ~ t<t_0} v(x_0, t)=:v_0
\end{equation}
exists but the derivatives $v_x(x,t)$ $v_t(x,t)$ blow up at the point $(x_0, t_0)$. Thus the solution $u(x,t;\varepsilon)$ to the Cauchy problem \eqref{burg1}, \eqref{burg2}, if exists, cannot be approximated by the inviscid solution. For the equation \eqref{burg0} the right asymptotic formula was found in \cite{ilin}. In the present section we will derive a suitable modification of this asymptotic formula and present some heuristic arguments justifying its validity. In the next section we will also give numerical evidences supporting our conjectures.

Let us begin with recollecting some basics from the method of characteristics for solving the inviscid equation \eqref{burg0}. For $t<t_0$ the solution to the inviscid equation can be represented in the following implicit form
\begin{equation}\label{sol1}
x=a(u) t + f(u)
\end{equation}
where the function $f(u)$ is inverse\footnote{If the initial data is not a globally monotone function then the representation \eqref{sol1} works on every interval of monotonicity.} to the initial data $v(x,0)$
\begin{equation}\label{sol2}
f\left( v(x,0)\right) \equiv x.
\end{equation}
Let $(x_0, t_0)$ be the point of gradient catastrophe of the solution. As above, denote
$$
v_0=v(x_0, t_0)
$$
the value of the solution at the point of catastrophe. The triple $(x_0, t_0, v_0)$ satisfies the following system of equations
\begin{eqnarray}\label{crit1}
&&
x_0=a_0 t_0 + f_0
\nonumber\\
&&
0= a_0' t_0 +f_0'
\\
&&
0=a_0'' t_0 +f_0''.
\nonumber
\end{eqnarray}
Here and below the following notations are used
$$
a_0=a(v_0),  \quad f_0=f(v_0), \quad a'_0=\left(\frac{da(v)}{dv}\right)_{v=v_0}, \quad a''_0=\left(\frac{d^2a(v)}{dv^2}\right)_{v=v_0}\quad \mbox{etc.}
$$
In the subsequent considerations we will always assume that
\begin{equation}\label{a0}
a_0'\neq 0.
\end{equation}
The genericity assumption
\begin{equation}\label{kap}
\kappa:=-\frac16 (a_0''' t_0 + f_0''')\neq 0
\end{equation}
ensures that the graph of the solution $v(x,t_0)$ has a non-degenerate inflection point at $x=x_0$. Such a solution will be called \emph{generic}. Locally a generic solution can be approximated by a cubic curve. For our subsequent considerations this well known statement can be presented in the following form (cf. \cite{univer1}).

\begin{lem} Near the point of gradient catastrophe a generic solution \eqref{sol1} to the inviscid equation \eqref{burg0} admits the following representation
\begin{eqnarray}\label{cubic1}
&&
v(x,t)=v_0+k^{1/3} \bar v (\bar x, \bar t) +{\mathcal O}\left( k^{2/3}\right), \quad k\to 0, \quad t<t_0
\\
&&\label{cubic2}
\bar x =\frac{x-x_0 -a_0(t-t_0)}{k}, \quad \bar t=\frac{t-t_0}{k^{2/3}}
\end{eqnarray}
where the function $\bar v(\bar x, \bar t)$ for $\bar t<0$ is defined as the (unique) root of the cubic equation
\begin{equation}\label{cubic3}
\bar x = a_0' \bar v\, \bar t -\kappa\, \bar v^3.
\end{equation}
\end{lem}

{\it Proof} can be easily obtained by substituting \eqref{cubic1}, \eqref{cubic2} into implicit equation \eqref{sol1} of the method of characteristics and then expanding with respect to the small parameter $k^{1/3}$. Observe that uniqueness of the root of the cubic equation \eqref{cubic3} for $\bar t<0$ is ensured by the condition
\begin{equation}\label{nerav}
a_0' \kappa >0
\end{equation}
valid due to a monotone decrease of the superposition $a\left( v(x, t_0)\right)$.

\begin{rmk} Observe that the cubic equation \eqref{cubic3} has a unique root also for $t>t_0$ provided validity of the inequality
\begin{equation}\label{cusp1}
\frac{|\bar x|}{\bar t^{3/2}} >\frac2{3\sqrt{3}} \left( \frac{{a_0'}^3}{\kappa}\right)^{1/2}.
\end{equation}
From this observation it is easy to derive existence and uniqueness of the solution $v(x,t)$ to \eqref{burg0} also for sufficiently small $t-t_0>0$ away from a cuspidal neighborhood
\begin{equation}\label{cusp2}
\frac{|x-x_0 -a_0(t-t_0)|}{(t-t_0)^{3/2}} < C\quad \mbox{\rm for some positive constant}\quad C
\end{equation}
of the point of catastrophe.
\end{rmk}

We are now in a position to formulate the main statement of the present paper.

\begin{conj}\label{con3} Let $v(x,t)$ be the solution to the inviscid equation \eqref{burg0} with a smooth monotone initial data $v(x,0)$ defined on  $\mathbb R\times [0, t_0)$ having a gradient catastrophe at the point $(x_0, t_0)$ satisfying \eqref{a0} and \eqref{nerav}. Assume the smooth function $b(u)$ to be such that
\begin{equation}\label{b0}
b_0:=b(v_0) >0.
\end{equation}
Then

\noindent 1) for sufficiently small $\varepsilon >0$ there exists a unique solution $u(x,t; \varepsilon)$ to the generalized Burgers equation \eqref{burg1} with the same $\varepsilon$-independent initial condition
$$
u(x,0;\varepsilon)=v(x,0), \quad x\in\mathbb R
$$
defined on $\mathbb R\times \left[0, t_0+\delta(\varepsilon)\right)$ for some sufficiently small $\delta(\varepsilon)>0$;

\noindent 2) away from a cuspidal neighborhood of the point of catastrophe the solution $u(x,t;\varepsilon)$ can be approximated by the inviscid solution $v(x,t)$
$$
|u(x,t;\varepsilon)-v(x,t)|={\mathcal O}(\varepsilon).
$$

\noindent For arbitrary $X$, $T$ there exists the limit
\begin{equation}\label{main1}
\lim_{\varepsilon\to 0+} \frac{u\left( x_0+a_0\beta\, \varepsilon^{1/2} T +\alpha \,\varepsilon^{3/4}, t_0+\beta\, \varepsilon^{1/2} T\right)-v_0}{\gamma\, \varepsilon^{1/4}}=:U(X,T)
\end{equation}
where
\begin{equation}\label{consts}
\alpha=\left(\frac{\kappa\, b_0^3}{{a_0'}^3}\right)^{1/4}, \quad \beta=\left( \frac{\kappa\, b_0}{{a_0'}^3 }\right)^{1/2}, \quad \gamma=\left( \frac{b_0}{\kappa\, a_0'}\right)^{1/4}.
\end{equation}
Moreover, the limit does not depend on the choice of solution neither on the choice of the $\varepsilon$-terms in the generalized Burgers equation \eqref{burg1}. It is given by the logarithmic derivative of the Pearcey function
\begin{equation}\label{main2}
U(X,T)=-2 \frac{\partial}{\partial X} \log \int_{-\infty}^\infty e^{-\frac18 \left( z^4 -2 z^2 T +4 z\, X\right)}dz.
\end{equation}
\end{conj}

A somewhat stronger version of the last statement of the Main Conjecture can be given in the form of the following asymptotic formula
\begin{equation}\label{main3}
u(x,t; \varepsilon) = v_0 +\gamma\, \varepsilon^{1/4} U\left( \frac{x-x_0-a_0(t-t_0)}{\alpha\, \varepsilon^{3/4}}, \frac{t-t_0}{\beta\, \varepsilon^{1/2}}\right) +{\mathcal O}\left( \varepsilon^{1/2}\right)
\end{equation}
expected to be true on some neighborhood of the catastrophe point. For the particular case $b(u)\equiv 1$, $c(u)\equiv 0$ the asymptotic formula \eqref{main3} coincides with the one obtained by A.M.~Il'in (see in \cite{ilin}).

Let us add few heuristic motivations of the Main Conjecture. First, let us consider the small time behavior of the solution $u(x,t;\epsilon)$. As the function $v(x,t)$ satisfies \eqref{burg1} modulo terms of order $\varepsilon$, one can seek the solution to the generalized Burgers equation in the form of a perturbative expansion
$$
u(x,t; \epsilon) =v(x,t) +\varepsilon\, v^{(1)} (x,t) +\varepsilon^2 v^{(2)}(x,t)+\dots
$$
The terms of the expansion have to be determined from linear inhomogeneous equations (see details in \cite{ilin}). For example, the first correction can be found from the following PDE
$$
v^{(1)}_t +\left( a(v) v^{(1)}\right)_x = b(v) v_{xx} + c(v) v_x^2.
$$
Instead, one can apply the method of the so-called \emph{quasitriviality transformations} \cite{DZ}, \cite{lz} finding a universal substitution
\begin{equation}\label{uquasi1}
v\mapsto u=v +\sum_{k\geq 1} \varepsilon^k \frac{f_k(v; v_x, v_{xx}, \dots, v^{(4k-2)}, \log | v_x|)}{v_x^{3k-2}}
\end{equation}
transforming any monotone solution of the inviscid equation \eqref{burg0} to a formal asymptotic solution to the perturbed equation \eqref{burg1}. Here $f_k(v; v_x, v_{xx}, \dots, v^{(4k-2)}, \log | v_x|)$ are some polynomials in the variables $v_x$, $v_{xx}$, \dots, $v^{(4k-2)}$, $\log | v_x|$ with coefficients that are smooth functions of $v$. They satisfy the following homogeneity condition
\begin{equation}\label{uquasi2}
f_k\left(v; \lambda\,v_x, \lambda^2v_{xx}, \dots, \lambda^{4k-2}v^{(4k-2)}, \log | v_x|\right)=\lambda^{4k-2}f_k(v; v_x, v_{xx}, \dots, v^{(4k-2)}, \log | v_x|), \quad k\geq 1
\end{equation}
for any $\lambda\neq 0$. Advantage of the perturbative expansion written in the form \eqref{uquasi1} is the \emph{locality} principle: changing the unperturbed solution within a small neighborhood of a point $(x^*, t^*)$ does not change the value of the perturbed solution outside the same neighborhood of the point.

For convenience of the reader let us explain the computational algorithm for derivation of the perturbative expansion \eqref{uquasi1}. For simplicity let us consider a perturbed equation of the form
\begin{equation}\label{pert1}
u_t+u\, u_x =\varepsilon\, \Phi(u; u_x, u_{xx}, \dots)
\end{equation}
where $\Phi(u; u_x, u_{xx}, \dots)$ is a smooth function of its variable polynomial in jets $u_x$, $u_{xx}$ etc. We will rewrite \eqref{pert1} as an equation for the function $x(u,t)$ inverse to $u(x,t)$:
\begin{equation}\label{pert2}
x_t =u-\varepsilon \, x_u\Phi\left( u; \frac1{x_u}, -\frac{x_{uu}}{x_u^3}, \dots\right).
\end{equation}
The clue is in the following statement (see \cite{lz}) describing the perturbative solution to \eqref{pert2}.

\begin{lem} Define the function $\Psi(u; x_u, x_{uu}, \dots)$ by the formula
\begin{equation}\label{pert3}
\Psi(u; x_u, x_{uu}, \dots)= \int x_u \Phi\left( u; \frac1{x_u}, -\frac{x_{uu}}{x_u^3}, \dots\right)\, dx_u.
\end{equation}
Then the function
\begin{equation}\label{pert31}
x(u, t) =x^{(0)}(u, t) -\varepsilon\, x^{(1)}(u,t)
\end{equation}
such that
\begin{equation}\label{pert4}
x^{(0)}_t=u, \quad x^{(1)}=\Psi\left(u;  x^{(0)}_u, x^{(0)}_{uu}, \dots\right)
\end{equation}
satisfies the perturbed equation \eqref{pert2} modulo terms of order $\varepsilon^2$.
\end{lem}

{\it Proof} immediately follows from independence from $t$ of higher $u$-derivatives of $x^{(0)}$:
$$
\frac{\partial}{\partial t} \frac{\partial^m x^{(0)}}{\partial u^m} = \frac{\partial^m }{\partial u^m}\frac{\partial x^{(0)}}{\partial t}=\delta_{m,1}\quad \mbox{for}\quad m\geq 1.
$$

\medskip

Inverting the series \eqref{pert31} one arrives at the needed algorithm.

\begin{cor}
Let $v=v(x,t)$ be a solution to the PDE
$$
v_t+v\, v_x=0
$$
satisfying $v_x\neq0$. Then the function
\begin{equation}\label{pert5}
u=v +\varepsilon\, v_x \Psi\left(v; \frac1{v_x}, -\frac{v_{xx}}{v_x^3}, \dots\right)
\end{equation}
satisfies the perturbed equation \eqref{pert1} modulo terms of order $\varepsilon^2$.
\end{cor}

For the particular case of the generalized Burgers equation \eqref{burg1} the first terms of the quasitriviality expansion read
\begin{equation}\label{quasi}
u=v -\varepsilon \left[ \frac{b}{a'} \frac{v_{xx}}{v_x} + \frac{c\, a' -b\, a''}{{a'}^2} v_x \log |v_x|\right]+{\mathcal O}\left(\varepsilon^2\right).
\end{equation}
It would be interesting to rigorously justify that, for sufficiently small $\varepsilon$  the above mentioned algorithm produces the asymptotic expansion of an actual solution to the generalized Burgers equation.

Let us now consider the solution to \eqref{burg1} in a neighborhood of the point of catastrophe. After a change of variables in the equation \eqref{burg1}
\begin{eqnarray}\label{scale1}
&&
x-x_0 -a_0(t-t_0)= \varepsilon^{3/4} \bar x
\nonumber\\
&&
t-t_0 =\varepsilon^{1/2} \bar t
\\
&&
u-v_0 = \varepsilon^{1/4} \bar u
\nonumber
\end{eqnarray}
one arrives at the equation
\begin{equation}\label{eq1}
\bar u_{\bar t} +a_0' \bar u\, \bar u_{\bar x} = b_0 \bar u_{\bar x\bar x} +{\mathcal O}\left( \varepsilon^{1/4}\right).
\end{equation}
Another substitution
\begin{equation}\label{abc}
\bar x=\alpha\, X, \quad \bar t=\beta\, T, \quad \bar u=\gamma\, U
\end{equation}
reduces the leading term of \eqref{eq1} to the standard form of the Burgers equation
$$
U_T+U\, U_X = U_{XX}
$$
provided the constants $\alpha$, $\beta$, $\gamma$ satisfy the constraints
\begin{equation}\label{ur1}
a_0' \frac{\beta\, \gamma}{\alpha}=1, \quad b_0\frac{\beta}{\alpha^2}=1.
\end{equation}
The Burgers equation can be solved by the Cole--Hopf substitution
$$
U(X,T) =-2 \frac{\partial}{\partial X}\log W(X,T)
$$
where $W=W(X,T)$ solves the heat equation
$$
W_T=W_{XX}.
$$
The Pearcey function
$$
W(X,T)=\int_{-\infty}^\infty e^{-\frac18 \left( z^4 -2 z^2 T +4 z\, X\right)}dz
$$
clearly satisfies the heat equation. Let us check that, using this function in the substitution
$$
\bar u=-2\gamma\frac{\partial}{\partial X} W(X,T)
$$
one arrives at the correct asymptotic expression of the function $\bar u$ near the point of catastrophe
\begin{equation}\label{cubic4}
\bar x = a_0' \bar u\, \bar t -\kappa \bar u^3 +{\mathcal O}\left( \varepsilon^{1/4}\right)
\end{equation}
(cf. eq. \eqref{cubic3} above). Indeed,
rescaling the integration variable
$$
\zeta=\varepsilon^{1/4} z
$$
we rewrite the expression for $\bar u$ in the form
$$
\bar u =-2 \alpha\, \gamma\, \varepsilon^{3/4} \frac{\partial}{\partial x} \log\int_{-\infty}^\infty e^{-\frac{S(\zeta; x,t)}{\varepsilon}}d\zeta
$$
where
\begin{equation}\label{action}
S(\zeta; x, t)=\frac18 \left( \zeta^4 -2 \zeta^2 \frac{t-t_0}{\beta} +4 \zeta\, \frac{x-x_0-a_0(t-t_0)}{\alpha}\right).
\end{equation}
For $t<t_0$ the phase function has a unique minimum at the point $\zeta_0=\zeta_0(x,t)$ determined by the cubic equation
\begin{equation}\label{crit}
 x-x_0-a_0(t-t_0)= \frac{\alpha}{\beta} ( t-t_0) \, \zeta_0 - \alpha \zeta_0^3.
\end{equation}
Applying the Laplace formula to the Pearcey integral
$$
\int_{-\infty}^\infty e^{-\frac{S(\zeta; x,t)}{\varepsilon}}d\zeta=\frac{2\sqrt{\pi\, \varepsilon}}{\sqrt{3\zeta_0^2 -\frac{t-t_0}{\beta}}}e^{-\frac{S(\zeta_0; x,t)}{\varepsilon}}\left( 1 +{\mathcal O}\left( \varepsilon\right)\right)
$$
and using the obvious formula
$$
\frac{\partial S\left( \zeta_0(x,t); x, t\right)}{\partial x}=\frac{\zeta_0(x,t)}{2\alpha}
$$
one arrives at the following expansion
$$
\bar u = \gamma\, \varepsilon^{-1/4} \zeta_0 \left( 1 +{\mathcal O}\left( \varepsilon\right)\right).
$$
Substituting into the cubic equation \eqref{crit} yields \eqref{cubic4} provided the constants $\alpha$, $\beta$, $\gamma$ satisfy one more constraint
\begin{equation}\label{ur2}
\frac{\alpha}{\gamma^3}=\kappa.
\end{equation}
Together with the constraints \eqref{ur1} this gives \eqref{consts}.

\vspace{\baselineskip}
\section*{2. Solving numerically the generalized Burgers equation. Comparison with the asymptotic formula.}
In order to test the numerical algorithms we will begin with the standard Burgers equation. First, let us consider the Cauchy problem for the inviscid equation
\begin{equation}\label{eq_DE1}
u_t+u u_x=0
\end{equation}
\begin{equation*}
u(x,0)=F(x)
\end{equation*}

At the point of catastrophe
one has
\begin{equation}\label{eq_DE3}
x_0=a_0+F(a_0)t_0, \quad t_0=-\dfrac{1}{F'(a_0)}, \quad u_0=F(a_0), \quad F''(a_0)=0
\end{equation}
(cf. eqs. \eqref{crit1} above).
For the particular choice of the initial data $F(x)=\dfrac{1}{1+x^2}$ the point of the catastrophe can be located as follows
\begin{equation}\label{eq_DE4}
x_0=\sqrt{3}, \quad t_0=\dfrac{8\sqrt{3}}{9}, \quad u_0=\dfrac{3}{4}.
\end{equation}

\begin{figure}[H]
  \centering
\includegraphics[scale=1]{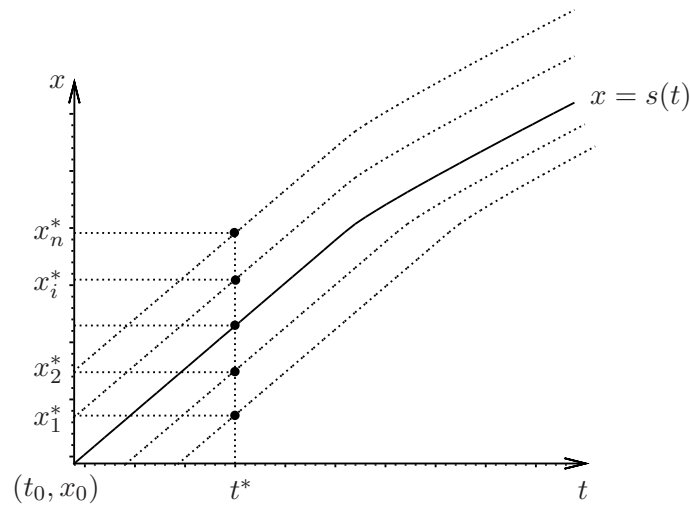}
  \caption{Shock front.}
  \label{fig1}
\end{figure}

For $t> t_0$ the solution to the Cauchy problem is close to a discontinuous one. Indeed,
it is well known (see, e.g., \cite{Whitham}) that the limit at $\epsilon\to 0$ of a smooth solution to the Burgers equation
\begin{equation}\label{eq_DE11}
u_t+u u_x=\varepsilon u_{xx}
\end{equation}
is described by a discontinuous function on the $(x,t)$-plane. The curve of discontinuity $x=s(t)$ of the limiting function is called \emph{shock front} (the solid line on Fig.~\ref{fig1}). We will be computing the numerical solution to the Cauchy problem in a neighborhood of the shock front and comparing it with the Il'in asymptotic formula. Let us explain the algorithm used for determination of the shock front.

Fixing a point $t=t^*$ we will select an array of values $\{x_i^*\}$ in some neighborhood of the curve $x=s(t)$. We will evaluate the function $u=u^*_i$ at the points $(t^*, x_i^*)$ with the help of the Il'in asymptotic formula using Maple for computation of the Pearcey function.

In order to determine the shock front (see \cite{Whitham}) let us use the Rankin--Hugoniot conditions
\begin{equation}\label{eq_DE5}
\dfrac{dx}{dt}=\dfrac{1}{2}(F(a_1)+F(a_2)),
\end{equation}
where $a_1(t)$ and $a_2(t)$ are determined by the equations of characteristics
\begin{equation}\label{eq_DE6}
x(t)=a_1+F(a_1)t, \quad x(t)=a_2+F(a_2)t.
\end{equation}
Differentiating (\ref{eq_DE6}) in $t$ and taking into account (\ref{eq_DE5}) one arrives at a system of differential equations for the functions $x(t)$, $a_1(t)$, $a_2(t)$
\begin{equation}\label{eq_DE7}
\dfrac{dx}{dt}=\dfrac{1}{2}(F(a_1)+F(a_2))
\end{equation}
\begin{equation*}
\dfrac{d a_1}{dt}=\dfrac{1}{2}\dfrac{F(a_2)+F(a_1)}{1+F'(a_1)t}
\end{equation*}
\begin{equation*}
\dfrac{d a_2}{dt}=\dfrac{1}{2}\dfrac{F(a_1)-F(a_2)}{1+F'(a_2)t}
\end{equation*}
The initial data for these equations have the form
\begin{equation}\label{eq_DE8}
a_1(t_0)=a_0, \quad a_2(t_0)=a_0, \quad x(t_0)=x_0,
\end{equation}
where $x_0$, $t_0$, $u_0$ are determined by eqs. (\ref{eq_DE3}).

If the solution to the Cauchy problem (\ref{eq_DE7})--(\ref{eq_DE8}) can be written in an explicit analytic form then also the shock front can be computed explicitly. Otherwise the system (\ref{eq_DE7})--(\ref{eq_DE8}) can be solved numerically. Observe that at $t=t_0$ one arrives at an ambiguity of the form  $\dfrac{0}{0}$. It can be resolved with the help of asymptotic expansions of the functions  $x(t)$, $a_1(t)$, $a_2(t)$ near the point $t=t_0$. If $a_1(t)<a_0<a_2(t)$, $t>t_0$ then for the characteristics $a_1(t)$, $a_2(t)$ we have
\begin{equation}\label{eq_DE9}
a_1(t)=a_0-\left(\dfrac{2F'(a_0)^2}{F'''(a_0)}(t-t_0)\right)^{1/2},
\quad
a_2(t)=a_0+\left(\dfrac{2F'(a_0)^2}{F'''(a_0)}(t-t_0)\right)^{1/2}.
\end{equation}
The expansion of $x(t)$ near $t=t_0$ has the form
\begin{equation}\label{eq_DE10}
x=x_0+F(a_0)(t-t_0).
\end{equation}
So, for solving the Cauchy problem (\ref{eq_DE7})--(\ref{eq_DE8}) we will solve the system of differential equations (\ref{eq_DE7}) where we put $t=t_0+\Delta t$. Here $\Delta t$ is the time step.
We use the asymptotic values (\ref{eq_DE9}), (\ref{eq_DE10}) as the initial data, i.e.
\begin{equation*}
a_1(t_0+\Delta t)=a_0-\left(\dfrac{2F'(a_0)^2}{F'''(a_0)}\Delta t\right)^{1/2},
\quad
a_2(t+\Delta t)=a_0+\left(\dfrac{2F'(a_0)^2}{F'''(a_0)}\Delta t\right)^{1/2}
\end{equation*}
\begin{equation*}
x(t_0+\Delta t)=x_0+F(a_0)\Delta t.
\end{equation*}
In order to control the computation the following identity will be used (see e.g. \cite{Whitham})
\begin{equation*}
\dfrac{1}{2}(F_1(a_1)+F_2(a_2))(a_1-a_2)=\int_{a_2}^{a_1}F(a)da.
\end{equation*}


\subsection*{Finite element analysis}

For solving the standard Burgers equation \eqref{eq_DE11}
we will use the finite element method (see \cite{Fletcher1}, \cite{Fletcher2},  \cite{Mitchell_Wait}, \cite{Strang_Fix}, \cite{Hecht}) realized in the package FreeFem++ \cite{Hecht}. Since this package, strictly speaking, is not designed for solving spatially one-dimensional problems one can reformulate the original problem as a 2D one considering solutions depending on one space variable only. Let us assume that the 2D domain has the rectangular form
\begin{equation*}
\Omega=\{ (x,y): 0\leq x\leq L_x, 0\leq y\leq L_y\},
\end{equation*}
of the size $L_x \times L_y$ and $L_x \gg  L_y$.

We will impose the no-flux boundary conditions at  $y=0$, $y=L_y$ but no specific values of $u$ at $x=0$, $x=L_x$ assuming that boundary values of $u$ are fixed at some fictitious boundary of a wider region
\begin{equation}\label{eq_DE12}
\left.\dfrac{\partial u}{\partial n}\right|_{y=0, L_y}=0 \quad \left. \dfrac{d u}{d t}\right|_{x=0, L_x}=0.
\end{equation}
Here $n$ is the exterior normal to the boundary $\partial\Omega$, $d/dt=\partial / \partial t+u \partial / \partial x$.

In the numerical experiments we will use the following initial data
\begin{equation}\label{eq_DE13}
\left. u\right|_{t=0}=\dfrac{1}{1+x^2}
\end{equation}
For the time approximation the semi-explicit Euler scheme will be used. To this end we multiply the equation by a test function $\theta$ integrating the resulting expression over the domain $\Omega$
\begin{equation*}
\iint\limits_\Omega \left( \dfrac{u^{m+1}-u^m}{\tau} \theta + u^m u_x^{m+1} \theta \right) dx dy = \iint\limits_\Omega \varepsilon u_{xx}^{m+1} \theta\, dx dy
\end{equation*}
or, taking into account the boundary conditions
\begin{equation}\label{eq_DE14}
\iint\limits_\Omega \left( \dfrac{u^{m+1}-u^m}{\tau} \theta + u^m u_x^{m+1} \theta + \varepsilon u_{x}^{m+1} \theta_x \right) dx dy = \int\limits_{\partial \Omega} \theta \dfrac{\partial u}{\partial n} ds
\end{equation}

The problem (\ref{eq_DE14}) in the weak formulation along with the initial conditions (\ref{eq_DE13}) is solved by means of the FreeFem++ package.

\begin{figure}[H]
\centering
\includegraphics[scale=1]{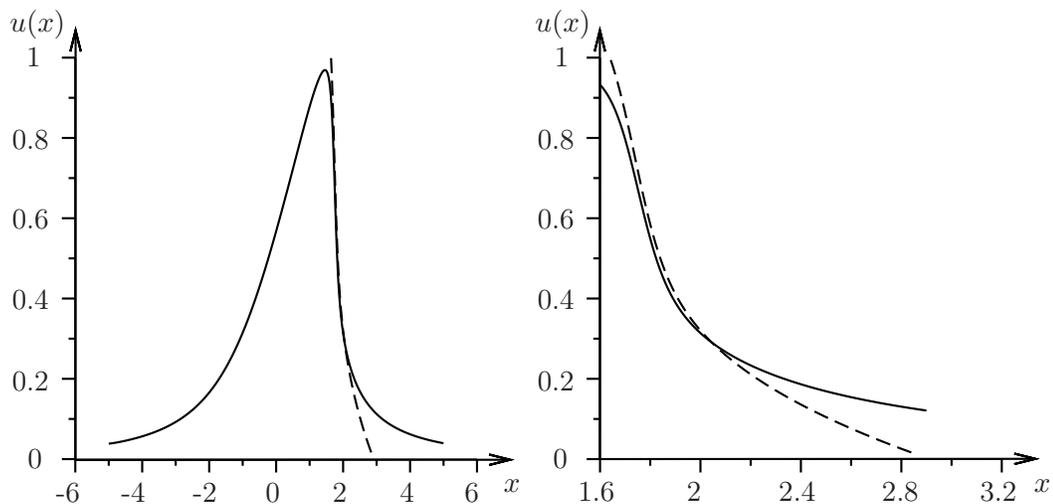}
\caption{Comparison of a numerical solution to Burgers equation with the Il'in's asymptotic formula for $\varepsilon=0.01$.}
  \label{fig_2}
\end{figure}

Comparison of the numerical solution with the Il'in's asymptotic formula near the shock front $x=s(t)$ for $t=1.54$, $\varepsilon=0.01$ is shown on Fig. \ref{fig_2}. The solid line shows the numerical solution obtained by the finite element method while the dashed one corresponds to the asymptotic solution \eqref{main3}.
On the right hand part of the figure the region near the catastrophe point
 $x_0$, $t_0$, $u_0$ (see (\ref{eq_DE4})) is zoomed in.

\begin{figure}[H]
\centering
\includegraphics[scale=1]{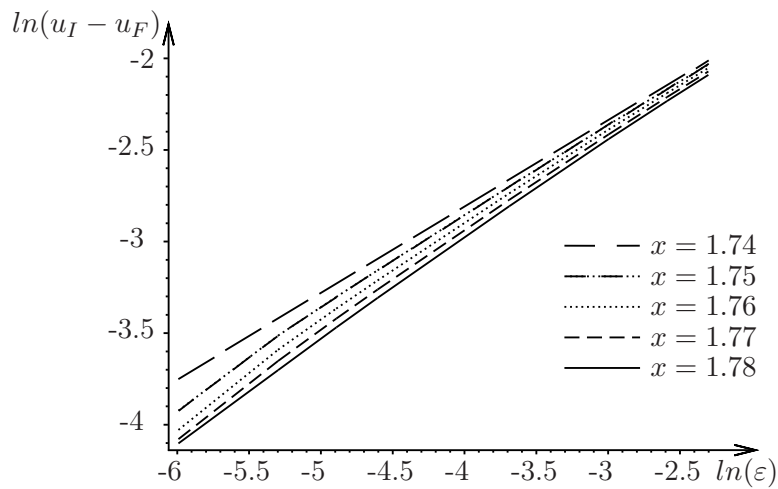}
\caption{Numerical estimate of the truncation error in the asymptotic formula.}
  \label{fig_3}
\end{figure}

On Fig. \ref{fig_3} the difference
between the asymptotic solution $u_I$ given by \eqref{main3} and the numerical solution
 $u_F$ is shown in the logarithmic scale. The evaluation of $u_I$ and $u_F$ is done for
 \begin{equation*}
t^*=1.54, \quad x^*=\{1.74, 1.75, 1.76, 1.77, 1.78\},
\end{equation*}
\begin{equation*}
\varepsilon=\{0.0025, 0.005, 0.0075, 0.01, 0.025, 0.05, 0.075, 0.1\}.
\end{equation*}
The average slope is $0.5175$ with the expected value $0.5$.

\begin{figure}[H]
\centering
\includegraphics[scale=1]{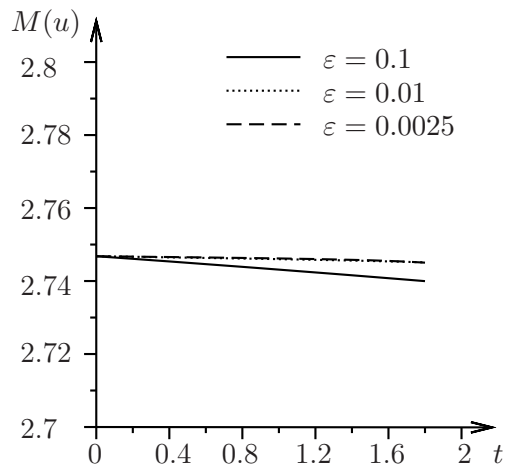}
\caption{Testing conservation of the total mass.}
  \label{fig_4}
\end{figure}

During the computation we control the total mass as function of time. With the boundary conditions under consideration the total mass is a conserved quantity. So the conservation of the total mass is a good test of the quality of numerical simulations.
The results for $\varepsilon=\{0.1, 0.01, 0.0025\}$ are shown on Fig.~\ref{fig_4}. On the interval $[0, 1.8]$ with $\varepsilon=0.1$ the relative error is $0.0024$, for $\varepsilon=0.01$ the relative error is $0.0006$ while for $\varepsilon=0.0025$ it drops to $0.0006$.

\subsection*{Generalized Burgers equation}

Let us now proceed to a particular example of the generalized Burgers equation \eqref{burg1}
\begin{equation}\label{eq_DE15}
u_t+u u_x=\varepsilon (uu_{x})_x.
\end{equation}
completed by the boundary conditions (\ref{eq_DE12}) and the initial data (\ref{eq_DE13}).
Like in the case of the standard Burgers equation (\ref{eq_DE11}) the semi-explicit Euler scheme will be used for the time approximation. The variational reformulation of the problem along with the boundary conditions reads
\begin{equation}\label{eq_DE16}
\iint\limits_\Omega \left( \dfrac{u^{m+1}-u^m}{\tau} \theta + u^m u_x^{m+1} \theta + \varepsilon u^m u_{x}^{m+1} \theta_x \right) dx dy = \int\limits_{\partial \Omega} \theta \dfrac{\partial u}{\partial n} ds
\end{equation}
The numerical solution of the problem (\ref{eq_DE15}), in the weak formulation, with the initial data (\ref{eq_DE13}) will be computed with the help of the FreeFem++ package.

\begin{figure}[H]
\centering
\includegraphics[scale=1]{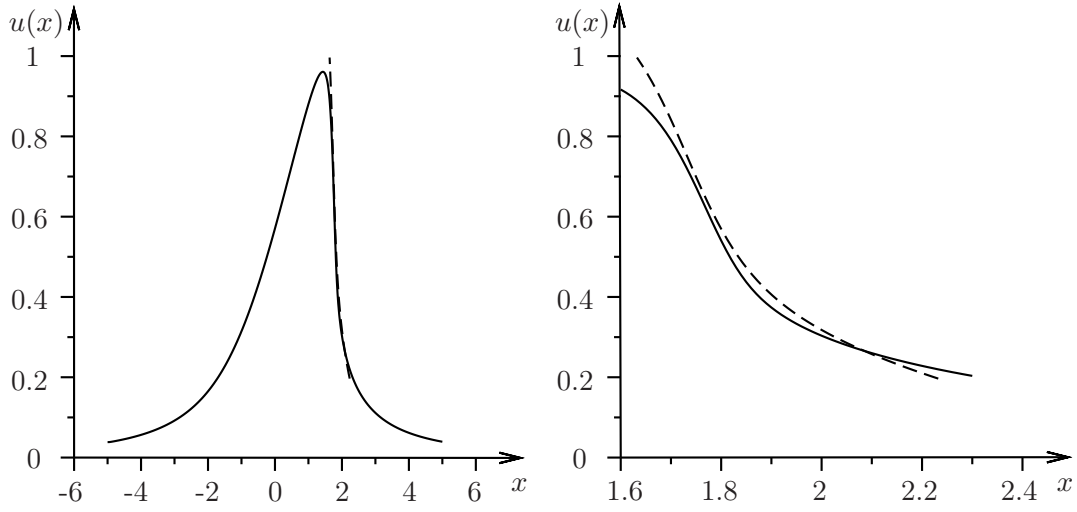}
\caption{Comparison of a numerical solution to the generalized Burgers equation with the asymptotic formula \eqref{main3} for $\varepsilon=0.01$.}
  \label{fig_5}
\end{figure}

Like above, let us compare the results of the numerical simulations with the predictions given by the asymptotic formula \eqref{main3}. On the Fig .~\ref{fig_5} the solid curve shows the numerical solution while the dashed one is the graph of the asymptotic solution \eqref{main3}. On the right hand part of the figure a neighborhood of the point of catastrophe $x_0$, $t_0$, $u_0$ is zoomed in.

\begin{figure}[H]
\centering
\includegraphics[scale=1]{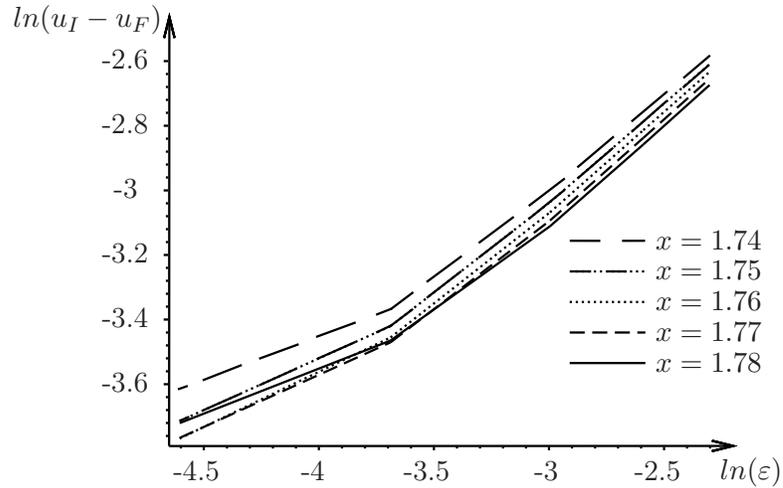}
\caption{Numerical estimate of the truncation error in the asymptotic formula \eqref{main3} for solutions to the generalized Burgers equation.}
  \label{fig_6}
\end{figure}

On Fig.~\ref{fig_6} the difference between the asymptotic formula $u_I$ and the numerical solution $u_F$ is shown in the logarithmic scale, for the following values of $(x, t, \varepsilon)$
\begin{equation*}
t^*=1.54, \quad x^*=\{1.74, 1.75, 1.76, 1.77, 1.78\},
\end{equation*}
\begin{equation*}
\varepsilon=\{0.01, 0.025, 0.05, 0.075, 0.1\}.
\end{equation*}
One can observe the average slope of $0.5221$ against the expected value $0.5$.

\begin{figure}[H]
\centering
\includegraphics[scale=1]{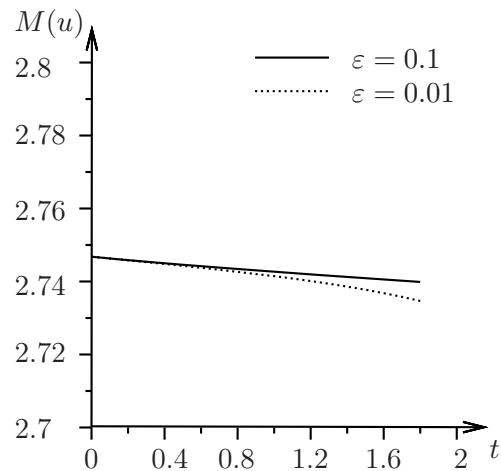}
\caption{Testing conservation of the total mass for the numerical solution to the generalized Burgers equation \eqref{eq_DE15}.}
  \label{fig_7}
\end{figure}

As above we used the conservation of the total mass valid for our particular case \eqref{eq_DE15} of the generalized Burgers equation as a test of validity of the numerical scheme. The results are shown on Fig.~\ref{fig_7} for the values $\varepsilon=\{0.1, 0.01\}$. On the interval  $[0, 1.8]$ for $\varepsilon=0.1$ the relative decay is $0.0025$, for $\varepsilon=0.01$ it drops to $0.0044$.

\end{document}